\documentstyle[preprint,aps]{revtex}
\begin{document}
\draft
\title{Non-RKKY oscillations of exchange coupling in magnetic multilayers}
\author{M. S. Ferreira \dag, J. d'Albuquerque e Castro\thanks
{Permanent address: Instituto de F\'{\i}sica, Universidade Federal
Fluminense, Niter\'oi, Rio de Janeiro, 24001-970,
Brazil}\ddag\hspace{0.1cm}, D. M. Edwards\dag \hspace{0.15cm} and
J. Mathon \ddag }
\address{\dag  Department of Mathematics, Imperial College, London,
SW7 2BZ, UK
\\
\ddag City University, London, EC1V 0HB, UK }
\date{\today}
\maketitle
\begin{abstract}
It is shown how the exchange coupling between two ferromagnetic planes
embedded in an infinite non-magnetic metal, regarded as a function of
the distance between the planes, may contain important components which
oscillate with periods not predicted by RKKY theory. The interesting
case of a FCC(110) structure with a Cu-like Fermi surface is discussed
in detail.
\end{abstract}
\pacs{75.50.Fr, 75.30.Et, 75.50.Rr}

Oscillatory exchange coupling between metallic magnetic
layers across a non-magnetic spacer has been intensively studied over the
last five years. One of the main issues in this area has
been the determination of the oscillation periods of the coupling
as a function of the spacer thickness.

The physical mechanisms which have been proposed for explaining this
phenomenon include the quantum well theory (QWT) of Edwards {\it et al.}
\cite{ed1,ed2,mt1} and an extension of the RKKY theory to the multilayer
geometry due to Bruno and Chappert \cite{bc1,bc2}. In the QWT electrons
propagating across the multilayer structure experience spin-dependent
potential wells whose depths depend on the exchange interaction in the
ferromagnetic layers. The oscillatory behaviour of the interlayer
coupling arises as a consequence of quantum interference effects inside
the wells, and bears a formal analogy to de Haas-van Alphen oscillations
\cite{ed1}. The existence of these quantum wells has been confirmed
experimentally by photoemission measurements \cite{Ortega}.

In the RKKY theory the oscillation periods are directly related to the
spacer Fermi surface (FS) and are given by the wave vectors ${\bf
q}_{z}$ perpendicular
to the layers that span the FS across those parts whose group velocities
are mutually antiparallel. There is a general belief that all
oscillation periods are given by the RKKY theory. Indeed, it has been
shown that in certain simple models \cite{ed1,ed2,mt1} the periods
predicted by the QWT coincide with the RKKY ones and are given by the
extremal dimensions of the spacer FS in the direction perpendicular to
the layers. In another case \cite{fcc111}, where the lattice lacks
reflection symmetry about a layer plane, the correspondence between the
quantum well and the RKKY periods is more subtle but still obtains.
Of course in the models mentioned above harmonics of the RKKY appear but
no new fundamental periods.
Furthermore, d'Albuquerque e Castro {\it et al.} \cite{nos1} showed
analytically for a very general model that RKKY theory holds in the
limit of very small exchange splitting in the ferromagnetic material.
However, van Schilfgaarde and Harrison found that real systems, such as
Fe/Cr, are not in this limit \cite{sh}, although they were convinced that the
oscillation periods were derivable from the RKKY. Here we show, however,
that under certain conditions, including those met in FCC(110)
multilayers with a Cu-like spacer FS, {\it oscillation periods exist
which are not predicted by RKKY}. These are in addition to the usual
RKKY periods.

A FCC(110) magnetic multilayer within the one-band nearest-neighbour
tight-binding model is an
interesting system because of the difficulty in determining analytically
the energies of the resonances and size
quantized states in such a structure \cite{harrison}. Here, however,
we approach the problem using the formalism of Ref. \cite{nos1}, which
gives the coupling in terms of the one-electron propagators. This
enables us to find the periods of oscillation analytically and to
evaluate the exchange coupling numerically.

We consider a multilayered system consisting of two parallel
ferromagnetic atomic planes embedded in an infinite
non-magnetic material. We label these two planes $0$ and $n$, so that
the number of atomic planes in the spacer layers is equal to $n-1$. It
has been shown recently that as far as the interlayer coupling as a
function of the spacer thickness is concerned, the thickness of the magnetic
layers affects the phase and amplitude of the oscillations, but not the
periods\cite{ze}.

Within the single band model, the expression for the exchange coupling
$J$, defined as the difference in the thermodynamical potential between
the ferromagnetic and antiferromagnetic configurations, is given by \cite{nos1}
\begin{equation}
J \,=\,- {1\over\pi} \,\sum_{{\bf q}_\parallel} \,
\int d\,\omega\,f(\omega)\,F({\bf q}_\parallel,\omega)\,,
\label{j1}
\end{equation}
where
\begin{equation}
F({\bf q}_\parallel,\omega) = Im\,ln\,\{\,{\,1 + 4 V^{\,2}_{ex}
\,G^{\uparrow}_{n 0}({\bf q}_\parallel,\omega)\,
G^{\downarrow}_{0 n}({\bf q}_\parallel,\omega)} \,\}\,,
\label{Fwq}
\end{equation}
and $G^{\sigma}_{n 0}({\bf q}_\parallel,\omega)$ is the off-diagonal
matrix element between planes $0$ and $n$ of the Green's function for an
electron with spin $\sigma$ in the ferromagnetic
configuration of the system, $f(\omega)$ is the Fermi function,
$V_{ex}$ is the exchange interaction in the ferromagnetic layers, and
the summation over ${\bf q}_\parallel$ is restricted to the
two-dimensional Brillouin zone (BZ). We assume for simplicity that the
site energies in
the ferromagnetic material are $\epsilon_{\uparrow,\downarrow} =
\epsilon_0 {\mp}V_{ex}$,
where  $\epsilon_0$ is the spacer on-site energy. In this situation,
the off-diagonal propagators in Eq.(\ref{Fwq})  can be written for each
${\bf q}_\parallel$ and $\omega$ as
\begin{equation}
G^{\uparrow ,\downarrow}_{n 0} ={\pm}{\tau_{\uparrow,\downarrow}\,
g_{n 0}\,(1-\tau_{\uparrow,\downarrow}\,g_{0 0}) \over
V_{ex}\,(1-\tau_{\uparrow,\downarrow}\,g_{n 0} \tau_{\uparrow,\downarrow}
\,g_{0 n})} \,,
\label{G}
\end{equation}
where $\tau_{\uparrow,\downarrow} = {\pm} V_{ex}(1 {\pm} V_{ex}g_{0 0})^{-1}$,
and $g_{n 0}$ is the matrix element of the bulk spacer Green's function.
A similar expression can be obtained for $G^{\uparrow,\downarrow}_{0 n}$.
In general, within the one-band model, $g_{0 n}$ is equal to
$g_{n 0}$, apart from a possible ${\bf q}_\parallel$-dependent phase factor.

It is clear that the behaviour of the coupling as a function of the spacer
thickness is related to the dependence of $g_{n 0}$ on $n$, which can be
determined as follows. The matrix element of $g$ between arbitrary
planes $l$ and $m$
is given by
\begin{equation}
g_{l m} = ({1 \over 2\pi}) \, \int_{-{\pi \over d}}^{\pi \over
d}\,\,dq_\perp \,{e^{-i q_\perp (l-m) d} \over \omega - E({\bf
q}_\parallel,q_\perp) + i\,0^+}
\label{integral}
\end{equation}
where $q_{\perp}$ is the wave vector perpendicular to the layers, $d$ is
the interplane distance, and $E({\bf q}_\parallel,q_\perp)$ describes
the bulk spacer band structure.
We evaluate the above expression for $g_{l m}$ with $l<m$ by integrating
around the boundary of the semi-infinite rectangle $-\pi/d\leq Re \,
q_\perp\leq\pi/d$ ; $Im \, q_\perp \geq 0$, in the complex $q_\perp$-plane.
This procedure is
completely general and can be applied, within the one-band model, to any
lattice structure and layer orientation, with hoppings to arbitrary number
of neighbours. For the FCC(110) case,
$E({\bf q}_\parallel,q_\perp) = \epsilon({\bf q}_\parallel) + 2 t_1({\bf
q}_\parallel)
cos(q_\perp d) + 2t_2({\bf q}_\parallel) cos(2q_\perp d)$, where
$\epsilon({\bf q}_\parallel) = - 2t_0\,cos(2q_x d)$,
$t_1({\bf q}_\parallel) = -4t_0\,cos(q_x d)cos(\sqrt 2 q_y d)$, and
$t_2({\bf q}_\parallel) = -t_0$. Here $q_x$ and $q_y$ are the components
of ${\bf q}_\parallel$, $-t_0$ is the
hopping between first nearest neighbour atoms, and the origin of
energy is chosen such that $\epsilon_0 = 0$. $t_1$ and $t_2$ are
the hoppings to first and second nearest planes, respectively. Using the
contour integration  described above, we obtain
\begin{equation}
g_{n 0}({\bf q}_\parallel,\omega) = A_1({\bf q}_\parallel,\omega)
e^{i q_1({\bf q}_\parallel,\omega) n d} + A_2({\bf
q}_\parallel,\omega) e^{i q_2({\bf q}_\parallel,\omega) n d},
\label{gn0}
\end{equation}
where $cos(q_j d) = -[\gamma+(-1)^j \sqrt{\gamma^2 + 8(\omega - \epsilon +
2t_2) /2t_2}]/4$, $A_j = [2i(cos(q_1 d) - cos(q_2 d))
(1-cos^2(q_j d))^{1/2}]^{-1}$, for $j=1,2$, and $\gamma=t_1/t_2$. Here
$q_\perp=\pm q_1$ and $q_\perp=\pm q_2$ are the roots of the
equation $E({\bf q}_\parallel,q_\perp) = \omega$. In the present case
$g_{0 n} = g_{n 0}$ and the above expression for $g_{n 0}$ can be
extended to continuous values of $n$.
Eq.(\ref{gn0}) shows that for values of ${\bf
q}_\parallel$ and $\omega$, for which $q_1$ and $q_2$ are real, $g_{n 0}$
oscillates with the superposition of two periods, $2\pi d/\vert q_1\vert$ and
$2\pi d/\vert q_2\vert$,
which are in general incommensurate. In those cases $g_{n 0}$ exhibits a
quasi-periodic dependence on $n$. As it is shown below, this fact may have
a striking effect on the coupling.

The function $F$ in Eq.(\ref{j1}) varies with $n$ through $g_{n 0}$ and
therefore exhibits the same quasi-periodic behaviour.
In order to deal with this quasi-periodic function, we make use of a
procedure analogous to
the one recently proposed \cite{ze} to investigate the dependence of the
coupling on the magnetic layer thickness. It consists in replacing $n$
in expression (\ref{gn0}) for $g_{n 0}$ by fictitious variables $n_1$
and $n_2$, which multiply
$q_1$ and $q_2$, respectively. The real physical situation corresponds to
$n_1 = n_2 = n$. The extended function $F(n_1 , n_2 )$ is then periodic in
each variable separately, and can be Fourier analysed in the usual way.
Thus we find
\begin{equation}
F = \sum_{m_1, m_2} \, C_{m_1 ,m_2 }\,e^{i(m_1 q_1 + m_2 q_2 )nd},
\label{F}
\end{equation}
where $C_{m_1 , m_2 }({\bf q}_\parallel,\omega)$ are the Fourier
coefficients, and $m_1$ and $m_2$ are
integers. Since $F$ is in fact a function of $g_{n 0}^2$, it follows
that $C_{m_1,m_2}=0$ unless $m_1+m_2$ is even. These coefficients
contain all the information about the electron potential in the
ferromagnetic layers. In particular, they depend on the magnitude of
$V_{ex}$ and vanish for
$V_{ex}=0$. In the RKKY limit, where $F$ is replaced by the leading
second-order term of its expansion in powers of $V_{ex}$, the Fourier
coefficients can be determined analytically. In this limit only six
coefficients appear, namely $C_{\pm 2 , 0}$, $C_{0, \pm 2 }$, and
$C_{\pm 1, \pm 1}$, whose values are given by $C_{2 , 0} = V_{ex}^2
A_1^2/2i$, $C_{0, 2} = V_{ex}^2 A_2^2/2i$, $C_{1 , 1} = V_{ex}^2 A_1
A_2/i$, and the property $C_{-m_1 , -m_2 } = C_{m_1 , m_2}^{*}$.

By inserting Eq.(\ref{F}) into Eq.(\ref{j1})
we find that, according to the usual stationary phase method
\cite{ed1,ed2,mt1,ze}, for sufficiently large values of $n$, the nonzero
contributions to the coupling come from $\omega$ equal to the Fermi
energy $E_{F}$ and ${\bf q}_\parallel$ in the
neighbourhood of those points at which the
argument of the exponential is stationary. They are given by the equation
\begin{equation}
m_1 {\bf \nabla}_{\parallel} q_{1}({\bf q}_{\parallel},E_F ) +
m_2 {\bf \nabla}_{\parallel} q_{2}({\bf q}_{\parallel},E_F ) = 0,
\label{grad}
\end{equation}
where ${\bf \nabla}_{\parallel}$ is the two-dimensional gradient in
${\bf q}_{\parallel}$ space. The position of the extremal points can be
determined exactly from the analytical expressions for $q_1$ and $q_2$.
The weight of the contribution to the coupling from each solution of
Eq.(\ref{grad}) depends on the magnitude of the corresponding
Fourier coefficient and partial derivatives of $\phi=m_1 q_1+m_2 q_2$
with respect to $q_x$, $q_y$, and $\omega$.

Clearly the surfaces $q_\perp=\pm q_{1}({\bf q}_{\parallel},E_F )$ and
$q_\perp=\pm q_{2}({\bf q}_{\parallel},E_F )$ map out the FS.
Fig. \ref{fs} shows a cross-section perpendicular to the atomic planes
of the FS for $E_F/2 t_0 = 1.64$ and $q_y
= \pi\sqrt{2}/4d$. Note that $q_1 > 0$ and $q_2 < 0$. Full lines
correspond to the surface $\pm q_1$ and dashed lines to $\pm q_2$. The
nature of the FS depends on $E_F$, and three distinct energy regions are
to be considered, namely $-12t_0 \leq E_F \leq -4t_0$,  $-4t_0 \leq E_F
\leq 0$, and $0 \leq E_F \leq 4t_0$. The two boundary values, $-4t_0$ and $0$,
correspond to those values of $E_F$ at which the FS first touches the
layer geometry BZ, and develops necks, respectively. As we show below,
the numbers of periods we obtain in the three regions are different.

It is interesting to
examine first the predictions for the periods in the RKKY limit, where
only six integers pairs $m_1\, m_2$ are to be considered in Eq.(\ref{grad}).
In the first
energy region we find only one period $\lambda_{2,0}^{\it a} = \pi
d/\vert q_1({\bf
q}_\parallel^{\it a}, E_F)\vert$, with ${\bf q}_{\parallel}^{\it a} = (0,0)$.
In the second region an additional period $\lambda_{1,1}^{\it b} =
2\pi d/\vert q_1({\bf q}_\parallel^{\it b}, E_F) + q_2({\bf
q}_\parallel^{\it b}, E_F) \vert$ appears, where ${\bf q}_\parallel^{\it b} =
(0,\pm \pi \sqrt{2}/4d)$. Finally, in the third region RKKY predicts four
periods, namely, $\lambda_{2,0}^{\it a}$, $\lambda_{2,0}^{\it c} = \pi
d/\vert q_1({\bf
q}_\parallel^{\it c}, E_F)\vert$,
$\lambda_{0,2}^{\it c} = \pi d/\vert q_2({\bf q}_\parallel^{\it c},
E_F)\vert$, and $\lambda_{1,1}^{\it c} =
2\pi d/\vert q_1({\bf q}_\parallel^{\it c}, E_F) + q_2({\bf
q}_\parallel^{\it c}, E_F) \vert = \lambda_{1,1}^{\it d} = 2\pi d/\vert
q_1({\bf
q}_\parallel^{\it d},E_F) + q_2({\bf
q}_\parallel^{\it d}, E_F) \vert$, with ${\bf q}_\parallel^{\it c} =
(\pm \pi/2 d,\pm \pi \sqrt{2}/4d)$ and ${\bf q}_\parallel^{\it d} =
(\pm \pi/2 d,0)$ . The Fourier coefficients associated
with the period $\lambda_{m_1,m_2}^{\it \alpha}$ are $C_{\pm m_1,\pm
m_2}({\bf q}_\parallel^{\it \alpha},E_F)$. These periods are
shown in Fig. \ref{periods} as functions of $E_F$. For the present
model we find that $q_1({\bf
q}_\parallel^{\it c},E_F)$ and $q_2({\bf q}_\parallel^{\it c},E_F)$,
which are represented in Fig. \ref{fs}, satisfy the relation $q_1({\bf
q}_\parallel^{\it c},E_F) - q_2({\bf q}_\parallel^{\it c},E_F) = \pi$.
Thus, the
oscillation periods $\lambda_{2,0}^{\it c}$ and $\lambda_{0,2}^{\it c}$
cannot be distinguished
just by looking at discrete integer values of $n$. We recall that in
FCC Cu $E_F$ lies in the third energy region with FS necks.
A quantitative description of the oscillation periods for Cu
can be obtained within the present framework by going beyond nearest
neighbours and using the tight-binding parameters of Halse \cite{halse}.
Then $\lambda_{1,1}^{\it c}$ and $\lambda_{1,1}^{\it d}$ become distinct
periods and, by taking into account the equivalence of
$\lambda_{2,0}^{\it c}$ and $\lambda_{0,2}^{\it c}$ for a discrete
lattice, we find exactly the four RKKY periods of Bruno and Chappert
\cite{bc1}.

\begin{figure}
\caption{Contributions to the coupling coming from the non-RKKY period
$\lambda_{3,1}^{\it c}$ (full line), and two RKKY periods
$\lambda_{2,0}^{\it a}$ (dashed line) and $\lambda_{2,0}^{\it c}$
(dot-dashed line) (see text).}
\label{stp}
\end{figure}
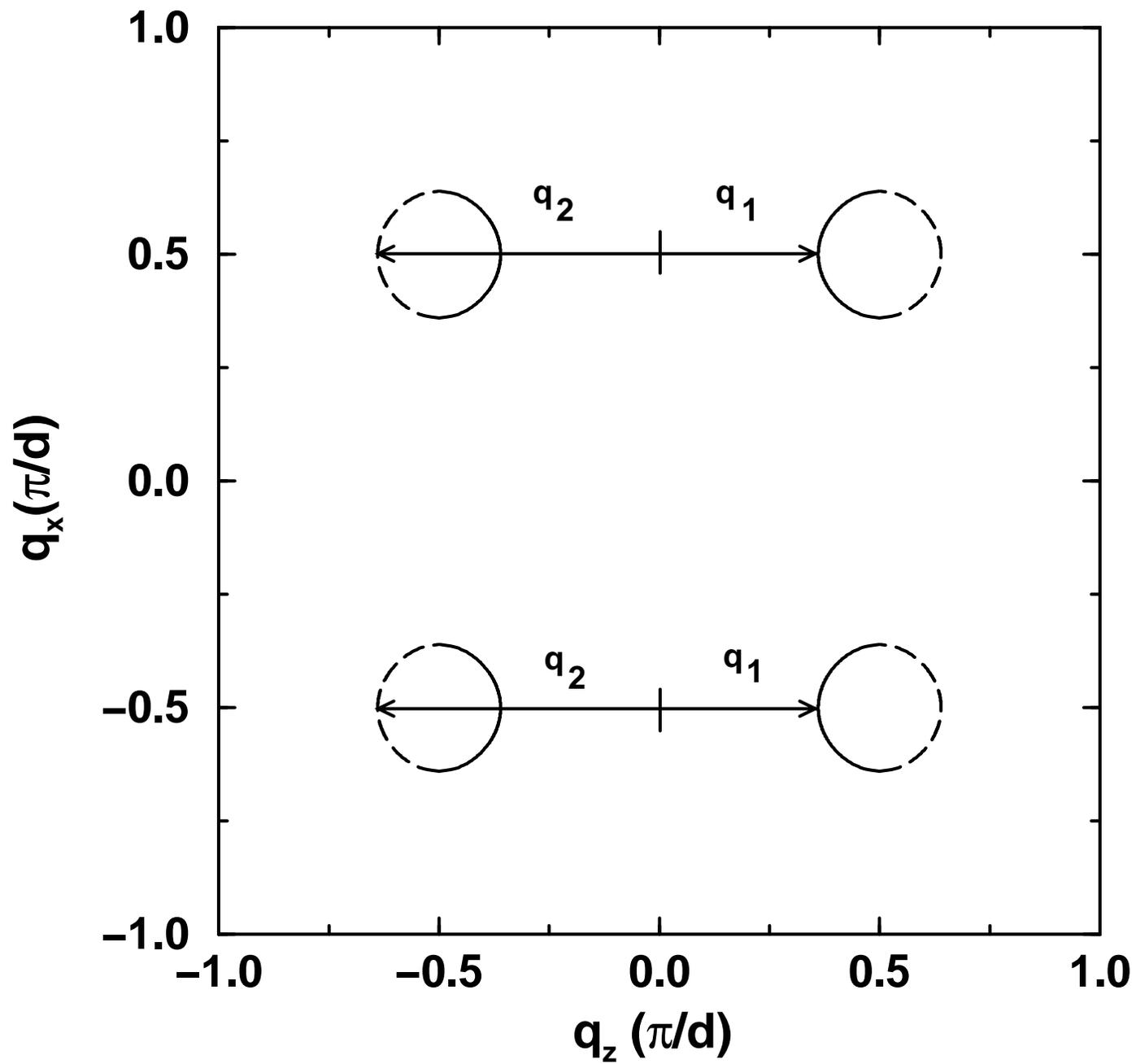
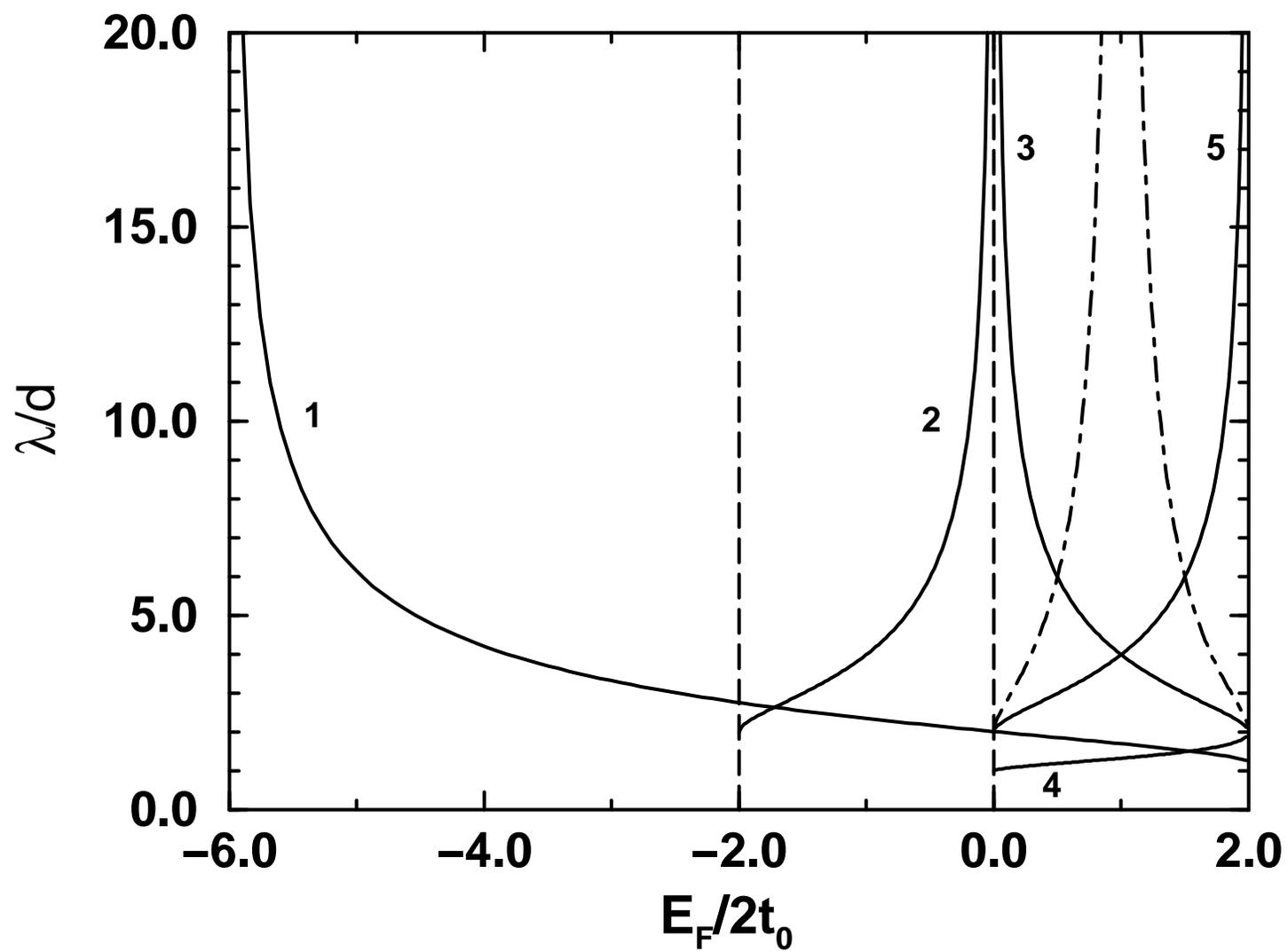
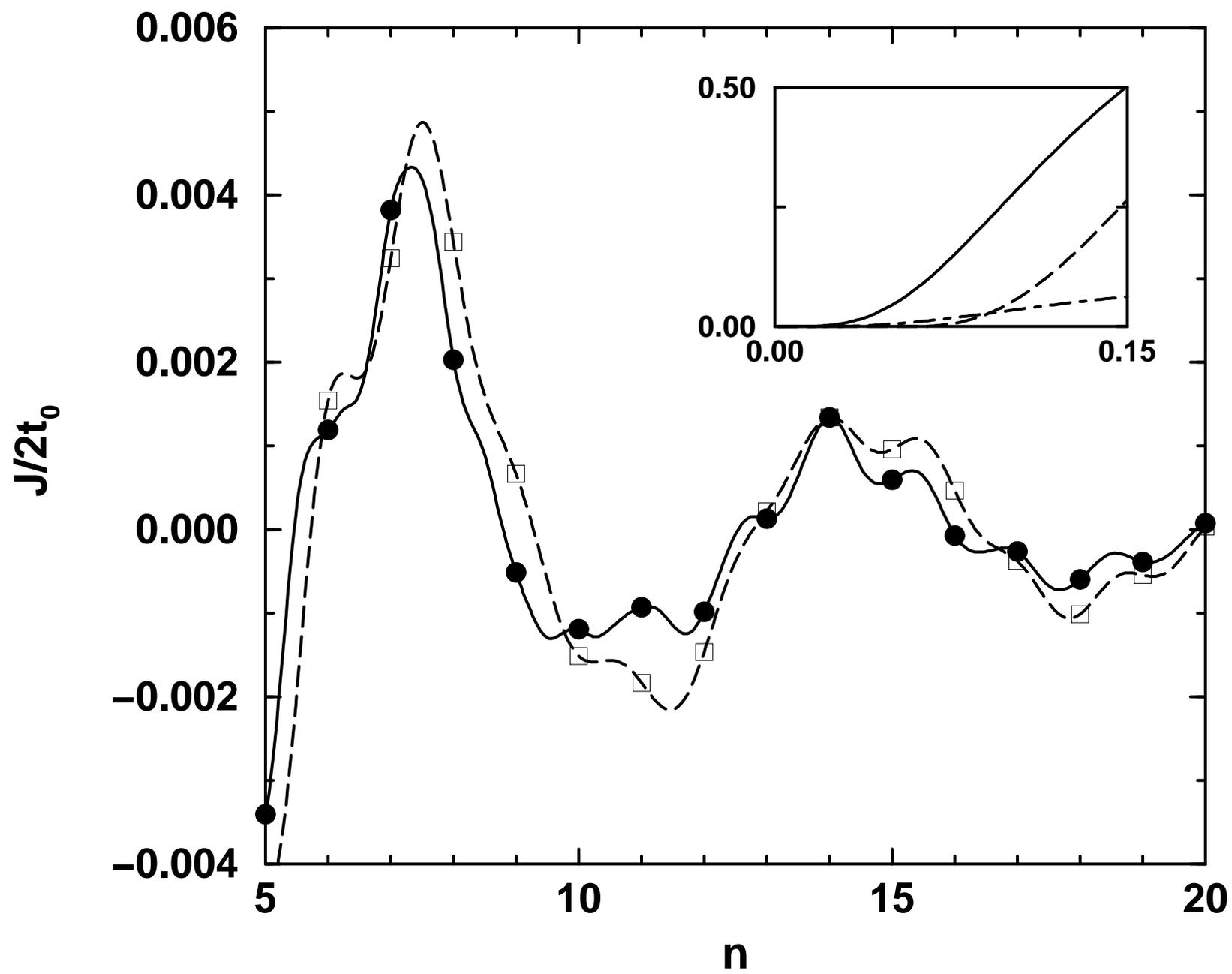
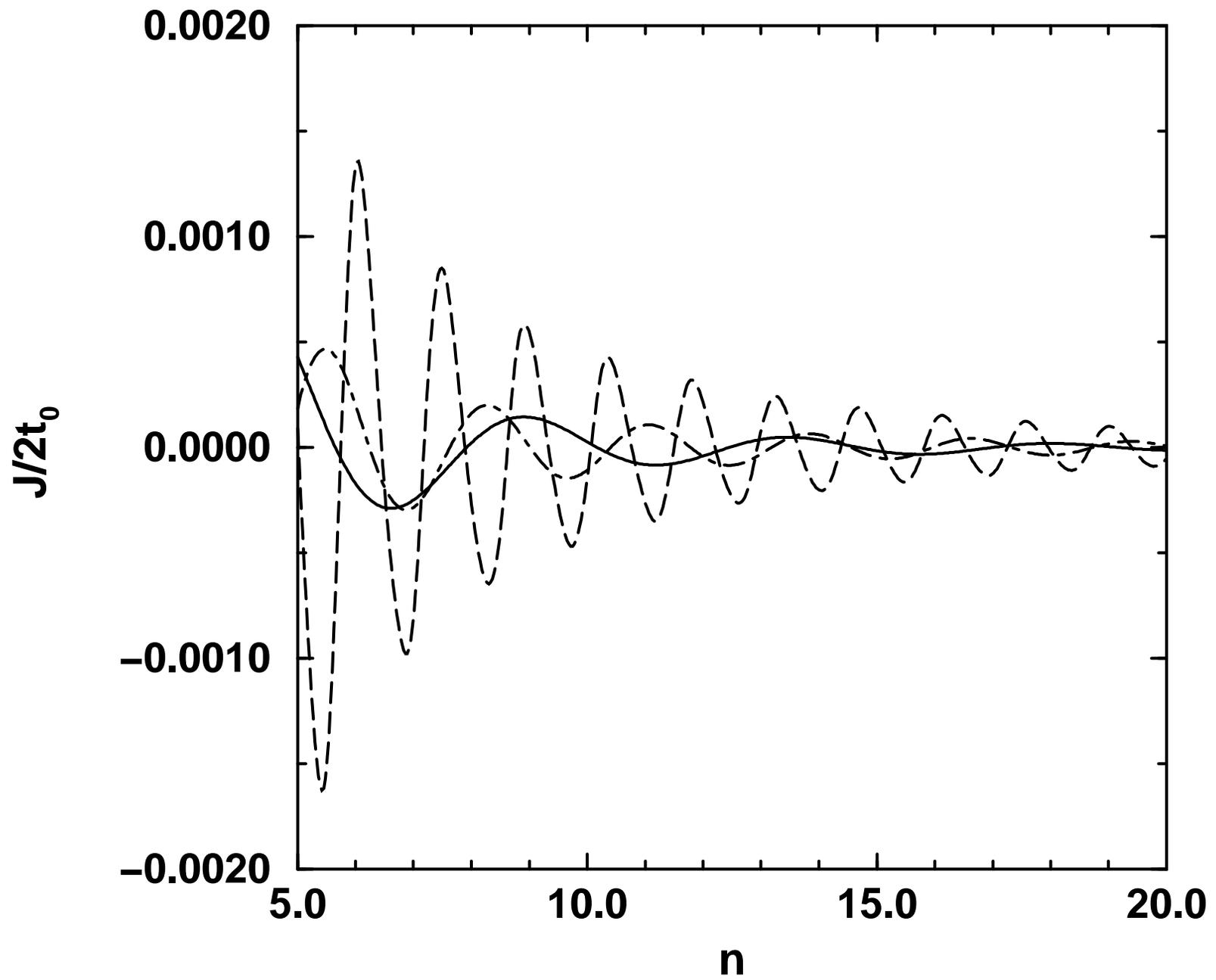

However, it follows from Eq.(\ref{grad}) that there are
contributions to the coupling with periods
other than those which arise in the RKKY limit discusssed above. In
fact, it is easy to show that for $E_F$ in the second and
third energy regions we find an infinite number of solutions,
corresponding to infinitely many values of $m_1$ and $m_2$.
Some of these solutions correspond merely to harmonics of the
fundamental RKKY periods but {\it some fundamentally new periods can
arise}. The simplest
and most interesting case actually occurs in the top energy region, where
${\bf \nabla}_{\parallel} q_{1}$ and $ {\bf \nabla}_{\parallel} q_{2}$
vanish simultaneously at ${\bf q}_\parallel^{\it c}$. Thus,
Eq.(\ref{grad}) is automatically satisfied for any values of $m_1$ and
$m_2$. The corresponding periods are $\lambda_{m_1,m_2}^{\it c} =
2\pi d/\vert m_1q_1 +
m_2q_2\vert$. Fig. \ref{coupling} exhibits the coupling $J$ at
temperature $T=2.0 \times
10^{-3} W/k_B$ as a function of the spacer thickness for $E_F/2 t_0 =
1.64$ and $V_{ex} = 0.15 W$, where $W= 16t_0$ is the spacer band-width.
We chose this value of $E_F$ so that an important new period
$\lambda_{3,1}^{\it c}$, plotted as a function of $E_F$ in
Fig.\ref{periods}, is well separated from the RKKY periods. The full line
in Fig.\ref{coupling} corresponds to the result obtained from
Eq.(\ref{j1}), and the dashed line
to the RKKY approximation scaled down by a factor of 8. In both cases $n$
was treated as a continuous variable, but the physical discrete values
are indicated. For this $E_F$ the long period dominates both curves
although with an amplitude differing by a large
factor, but the interesting fine structure is different and reflects
contributions beyond the fundamental RKKY periods.
To make this point explicit, we present in the inset the absolute value
of the ratio between some coefficients $C_{m_1,m_2}$
and $C_{1,1}$, the largest coefficient of a fundamental RKKY
period, as a function of $V_{ex}$ and for ${\bf q}_\parallel^{\it c}$.
As expected, for very small exchange splittings, the magnitudes of
higher order Fourier coefficients relative to that of the fundamental
RKKY one are negligible. However, they
rapidly increase with $V_{ex}$, and the ratio $\vert C_{m_1,m_2} \vert /
\vert C_{1,1} \vert$ becomes significant.
{}From the inset in Fig.\ref{coupling}, we see that for $V_{ex}=0.15 W$,
there are additional contributions to the coupling coming not just from
harmonics of the RKKY frequencies, which correspond to $m_1$ and $m_2$
even, but a very important one coming from a new period
$\lambda_{3,1}^{\it c} = 4.5 d$. In fact, the contribution from this new
period can be calculated separately using the stationary phase method
\cite{ed1,ed2,mt1,ze}. The result is shown in Fig.\ref{stp},
together with those corresponding to two of the fundamental RKKY
periods, namely,
$\lambda_{2,0}^{\it a}$ and $\lambda_{2,0}^{\it c}$. It is
clear that this new period is as important as the RKKY ones, except for
the dominant long period whose large amplitude is due to FS geometry
which causes some second derivatives of $\phi$ to vanish, making a
stationary phase evaluation of this contribution impossible.
The amplitude of the new contribution falls off as $1/n^2$ just like the
normal RKKY components and their harmonics \cite{ed1,ed2,mt1,bc1,bc2}.

The appearance of the non-RKKY periods is clearly related to the
spacer off-diagonal propagators oscillating as a function
of the spacer thickness with more than one period, for fixed energy
and ${\bf q}_\parallel$. As we have shown, such a behaviour can be
found even in the one-band model, for which the FS is simple and has
a single sheet. It can also be found in those cases in which the
spacer FS has more than one sheet. Thus we may expect the occurrence
of non-RKKY periods in the coupling through a non-magnetic transition
metal. Therefore, in those cases, the interpretation of the results in
terms of just the RKKY theory may be misleading.

In conclusion, we have shown that the relation between the oscillation
periods of the coupling and the spacer FS is more complex and subtler
than has been assumed so far, making possible the appearance of
non-RKKY periods. This is the central result in this communication, which
settles the long standing question about whether or not the quantum
well and the RKKY theories always give the same periods of oscillation.

We are grateful to EPSRC and Royal Society of UK, and CNPq of Brazil
for financial support. We also would like to acknowledge Dr. R. B. Muniz
and Dr. Murielle Villeret for useful discussions in the early stages of
this work.

\begin{figure}
\caption{Cross-section of the spacer Fermi surface perpendicular
to the layers for $E_F/2t_0=1.64$ (see text).
Full lines correspond to the surface $\pm q_1$ and
dashed lines to $\pm q_2$. The two vectors
$q_1({\bf q}_\parallel^{\it c},E_F)$ and
$q_2({\bf q}_\parallel^{\it c},E_F)$ are indicated.}
\label{fs}
\end{figure}
\begin{figure}
\caption{RKKY periods as a function of $E_F$. Curves 1,2,3,4, and 5
correspond to $\lambda_{2,0}^{\it a}$, $\lambda_{1,1}^{\it b}$,
$\lambda_{2,0}^{\it c}$, $\lambda_{0,2}^{\it c}$, and
$\lambda_{1,1}^{\it c} = \lambda_{1,1}^{\it d}$, respectively. The
dot-dashed curve is a new period $\lambda_{3,1}^{\it c}$.}
\label{periods}
\end{figure}
\begin{figure}
\caption{Exchange coupling as a function of the spacer thickness for
$E_F/2t_0 = 1.64$, $V_{ex} = 0.15 W$ and $k_B T = 2.0 \times 10^{-3}W$
(full line). The dashed line corresponds to the RKKY result scaled down
by a factor of 8. The inset shows the ratio $\vert C_{3,1} \vert /
\vert C_{1,1} \vert$ (full line), $\vert C_{4,2} \vert / \vert C_{1,1}
\vert$ (dashed line), and $\vert C_{4,0} \vert /
\vert C_{1,1} \vert$ (dot-dashed lined) as a function of $V_{ex}$, for
${\bf q}_\parallel^{\it c}$.}
\label{coupling}
\end{figure}
\end{document}